\documentclass[aps,prl,twocolumn,superscriptaddress,showpacs]{revtex4-1}
\usepackage{hyperref}
\usepackage{amsmath}
\usepackage{graphicx}

\begin{document}

\title{Pressure-Driven Orbital Selective Insulator to Metal Transition and Spin State Crossover in Cubic CoO}
\author{Li Huang}
\affiliation{ Beijing National Laboratory for Condensed Matter Physics, 
              and Institute of Physics, 
              Chinese Academy of Sciences, 
              Beijing 100190, 
              China }
\affiliation{ National Key Laboratory for Surface Physics and Chemistry, 
              P.O. Box 718-35, 
              Mianyang 621907, 
              Sichuan, 
              China }

\author{Yilin Wang}
\affiliation{ Beijing National Laboratory for Condensed Matter Physics, 
              and Institute of Physics, 
              Chinese Academy of Sciences, 
              Beijing 100190, 
              China }

\author{Xi Dai}
\affiliation{ Beijing National Laboratory for Condensed Matter Physics, 
              and Institute of Physics, 
              Chinese Academy of Sciences, 
              Beijing 100190, 
              China }

\date{\today}

\begin{abstract}
The metal-insulator and spin state transitions of CoO under high pressure are studied by using density 
functional theory combined with dynamical mean-field theory. Our calculations predict that the 
metal-insulator transition in CoO is a typical orbital selective Mott transition, where the $t_{2g}$ 
orbitals of Co $3d$ shell become metallic firstly around 60 GPa while the $e_g$ orbitals still remain 
insulating until 170 GPa. Further studies of the spin states of Co $3d$ shell reveal that the orbital 
selective Mott phase in the intermediate pressure regime is mainly stabilized by the high-spin state 
of the Co $3d$ shell and the transition from this phase to the full metallic state is driven by the 
high-spin to low-spin transition of the Co$^{2+}$ ions. Our results are in good agreement with the most 
recent transport and x-ray emission experiments under high pressure.
\end{abstract}


\maketitle

Although the Mott metal-insulator transition (MIT) has been studied extensively for decades, most of the works 
are focused on the single band Hubbard model, where the Mott transition is driven completely by the ratio of
the local Coulomb interaction and band width. While most of the Mott MITs in realistic materials\cite{imada:1039} 
involve more than one band where the transition is driven not only by the local Coulomb interaction but also 
by the distribution of the electrons among these bands. For example, the redistribution of the four electrons 
among three bands may lead to so-called orbital selective Mott transition (OSMT)\cite{anisimov:191,liebsch:226401,
koga:216402,koga:045128,de'medici:205124,medici:126401} in the $t_{2g}$ bands. On the other hand, in many systems 
the redistribution of the electrons among different bands is induced by the crossover in spin states, i.e. the 
high-spin (HS) to low-spin (LS) transition\cite{werner:126405}. Therefore in realistic materials (e.g. in $3d$ 
transition metal compounds) the Mott MIT and spin state crossovers are closely related to each other\cite{kunes:2008,
kunes:146402,shorikov:195101,lyub:085125,cohen:654,gavr:155112,mattila:196404}.

Recently the high pressure experiments on charge transfer insulator CoO revealed very interesting behaviors in both 
transport properties and x-ray emission spectroscopy (XES). The transport measurement\cite{atou:1281} indicated 
that with the increment of pressure there are two transitions in resistivity, one happens around 60 GPa and the 
other one takes place around 130 GPa. While the room temperature XES measurements\cite{rueff:s717,kurm:165127} on 
the similar sample show that the spin state of Co$^{2+}$ ions persist in the HS state all the way to 140 GPa, 
after which the crossover from HS to LS states happens. Hence the interplay between the HS-LS transition and the 
two-step Mott insulator transition becomes the key factor to understand the underlying physics in CoO.

The pressure-driven MIT and magnetic moment collapse in transition metal oxides have been studied by first principles 
calculations widely\cite{kunes:2008,kunes:146402,shorikov:195101,cohen:654,kas:195110,wdowik:1698}. As to CoO, its 
magnetic state transition under pressure was first discussed\cite{cohen:654} within the Stoner scenario by employing 
the local spin density approximation (LSDA) and generalized gradient approximation (GGA) approaches of density functional 
theory. A transition from HS to nonmagnetic metallic state was found around 88 GPa. Since these calculations did 
not take the correlation effects of the Co $3d$ shell into considerations, so the experimentally observed excitation 
gap for CoO\cite{kurm:165127} as large as 2.5 $\sim$ 2.6 eV was not reproduced completely. Recently, Zhang et al.\cite{zhang:155123} 
reinvestigated the pressure-driven magnetic phase transition in CoO by using LSDA + $U$ 
approach. The HS-LS transition is indeed obtained to be of $t^{5}_{2g}e^{2}_{g} \rightarrow t^{6}_{2g} e^{1}_{g}$ 
character, but the electronic structure transition is insulator to insulator scheme. This contradicts with the 
resistivity data under high pressure\cite{atou:1281}, which shows dramatic change in resistivity indicating the 
insulator to metal transition with pressure. 

In this letter, based on the local density approximation (LDA) combined with dynamical mean-field theory (DMFT)\cite{antoine:13,kotliar:865},
we have carried out theoretical calculations for cubic phase CoO at different volumes for the first time. Our 
calculations show that CoO under pressure is a typical system which has OSMT. At ambient pressure our calculation 
gives correct Mott insulator phase for CoO with energy gap being around 2.4 eV. At the first transition around 
60 GPa the $t_{2g}$ bands become metallic while the $e_{g}$ bands still remains insulating until the pressure 
reaches 170 GPa. Therefore in CoO the exotic orbital selective Mott phase (OSMP) with metallic $t_{2g}$ and insulating 
$e_g$ bands is stable in a quite large pressure window between 60 and 170 GPa. Our LDA+DMFT calculations also find 
that the Co$^{2+}$ ions remain in HS state during the first transition and the crossover to the LS state starts 
only after the second transition, which is in good agreement with both the resistivity\cite{atou:1281} and XES data
\cite{rueff:s717,kurm:165127} in CoO. 

The LDA+DMFT calculations\cite{korotin:91,trimarchi:135227,amadon:205112} in the present paper have been carried out
by using the pseudopotential plane-wave method with the {\scriptsize QUANTUM ESPRESSO} package\cite{qe2009} for the 
LDA part and continuous-time quantum Monte Carlo (CTQMC)\cite{werner:076405,gull:349} as the impurity solver for the 
DMFT part. The single particle LDA Hamiltonian is obtained by applying a projection onto atomic-centered symmetry-constrained 
Wannier function (WF) orbitals including all the Co $3d$ and oxygen $2p$ orbitals, which is described in details in 
Ref.\onlinecite{korotin:91}. That would correspond to a $8 \times 8$ $p-d$ Hamiltonian which is a minimal model required 
for a correct description of the electronic structure of CoO due to its charge transfer nature\cite{imada:1039}.
 
The LDA+DMFT calculations presented below have been done for crystal volumes corresponding to values 
of pressure up to 280 GPa. For simplicity, all first principles calculations were performed in nonmagnetic 
configuration for rocksalt-type crystal structure with lattice constant scaled to give a volume corresponding 
to applied pressure. The Coulomb interaction is taken into considerations merely among Co $3d$ orbitals. In 
the present work, we choose $U$ = 8.0 and $J$ = 0.9 eV, which are close to previous estimations\cite{zhang:155123,anis:943}.
We adopt the scheme proposed in reference \onlinecite{shorikov:195101} to deal with the double counting energy.
The effective impurity problem for the DMFT was solved by the CTQMC quantum impurity solver (hybridization 
expansion version) supplemented with recently developed orthogonal polynomial representation algorithm\cite{boe:075145}. 
The maximum entropy method\cite{jarrell:133} was used to perform analytical continuation to obtain the impurity spectral function of Co $3d$ states.
Calculations for all crystal volumes were preformed in paramagnetic state at the temperature of 290\ K.

\begin{figure}
\centering
\includegraphics[scale=0.56]{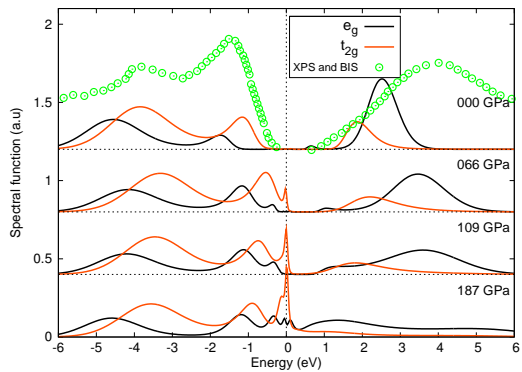}
\caption{(Color online) Single particle spectral function of Co $3d$ states vs pressure obtained in LDA+DMFT 
calculations at room temperature. The spectral function is obtained from imaginary-time green's function
$G(\tau)$ by using maximum entropy method\cite{jarrell:133}, and the results are cross-checked by using 
recently developed stochastic analytical continuation method\cite{beach}. The available x-ray photoelectron 
spectroscopy (XPS) and bremsstrahlung isochromat spectroscopy (BIS) experimental data\cite{kurm:165127} 
are drawn in this figure as a comparison.\label{fig:spectrum}}
\end{figure}
In Fig.\ref{fig:spectrum}, the evolution of single particle spectrum for Co $3d$ states upon compression
is shown. The momentum integrated  spectral function $A(\omega)$ under ambient pressure shows well defined 
insulating behavior for all $3d$ orbitals. However the energy gap for $e_g$ orbitals is slightly higher than 
that for $t_{2g}$ states indicating that the latter orbitals are closer to MIT than the former ones. At zero 
pressure, the calculated gap for $t_{2g}$ orbitals of about 2.1 eV and for $e_g$ orbitals of about 2.3 eV, 
which agree well with optical experimental value 2.6 eV\cite{kurm:165127}. The LDA+DMFT calculations made 
for small volume values corresponding to high pressures gave metallic state for CoO starting from 60 GPa in 
agreement with room temperature resistivity data\cite{atou:1281}. One can see that $t_{2g}$ orbitals become 
metallic whereas $e_g$ ones still remain insulating. This behavior reminds the OSMT scenario as discovered in 
ruthenates\cite{anisimov:191} at first. The spectral functions for $t_{2g}$ orbitals in Fig.\ref{fig:spectrum} 
for pressure values larger than 60 GPa become typical for strongly correlated metal close to MIT: well pronounced 
Hubbard bands and narrow quasiparticle peek. However, the $A(\omega)$ for $e_{g}$ bands remains insulating with 
Hubbard bands only but the gap size is strongly reduced comparing with that of ambient pressure. When the pressure 
exceeds about 170 GPa, the $e_g$ states undergo a insulator to metal transition. As is seen in Fig.\ref{fig:spectrum}, 
at 187 GPa the Mott gaps for both $3d$ orbitals disappear finally.

\begin{figure}
\centering
\includegraphics[scale=0.56]{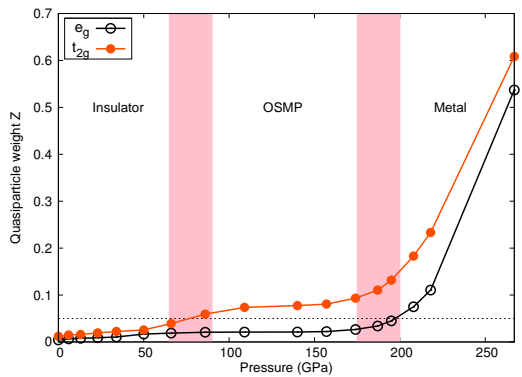}
\caption{(Color online) Quasiparticle weight Z of Co $3d$ states as a funtion of pressure. The transition 
zones are highlighted by pink vertical bars.\label{fig:quasi}}
\end{figure}
In order to reveal the nature of OSMT in Co $3d$ orbitals upon compression, we make further estimation for 
their quasiparticle weights by using the well-known equation\cite{antoine:13}:
$Z^{-1} = 1 - \frac{\partial}{\partial \omega} \text{Re}\Sigma(i\omega)|_{\omega = 0}$, 
where $\text{Re}\Sigma(i\omega)$ is the real part of impurity self-energy function at real frequency axis. 
In Fig.\ref{fig:quasi} the calculated quasiparticle weights for $t_{2g}$ and $e_{g}$ states as a function 
of pressure are shown. It is apparent that the phase diagram can be splitted into three different zones: 
(1) 0 GPa $<$ P $<$ 60 GPa. The quasiparticle weights for both the $e_{g}$ and $t_{2g}$ orbitals approach 
zero, and the system exhibits insulating behavior. (2) 60 GPa $<$ P $<$ 170 GPa. The quasiparticle weights 
for $t_{2g}$ states become considerable, whereas those for $e_g$ states remain very tiny. At this range of 
pressure, it gives an exotic OSMP. (3) P $>$ 170 GPa. The quasiparticle weights for both $t_{2g}$ and $e_{g}$ 
states show a dramatic increment with pressure, and the system goes into the fully metallic state. 

\begin{figure}
\centering
\includegraphics[scale=0.35]{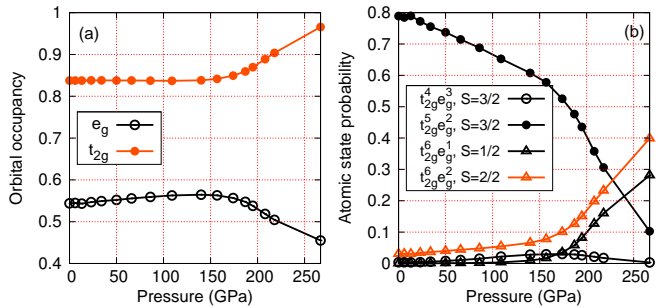}
\caption{(Color online)(a) The orbital occupancy of Co $3d$ states as a function of external pressure. (b)
The principal pressure-dependent atomic state probability of Co $3d$ states obtained by LDA+DMFT 
calculations.\label{fig:occ}}
\end{figure}
In Fig.\ref{fig:occ}(a) we show the evolution of Co $3d$ occupancies and the atomic state probability under 
compression. Due to the charge transfer from O $2p$ orbitals to Co $3d$ orbitals, the total $3d$ states 
occupation number is $\sim$ 7.2, which is slightly larger than the nominal value 7.0. At ambient pressure, 
the occupation numbers for $t_{2g}$ and $e_{g}$ orbitals are $n(e_{g}) = 0.54$ and $n(t_{2g}) = 0.84$, 
respectively. Those numbers agree very well with the HS state of Co$^{2+}$ ion in cubic crystal field with two 
electrons in $e_{g}$ states and five electrons in $t_{2g}$ states. Over the pressure range from 0 to 170 GPa, 
the occupation numbers for $3d$ orbitals basically remains unchanged. While when the pressure is larger 
than 170 GPa, the occupation numbers for $t_{2g}$ and $e_{g}$ states show a dramatic change. The $n(t_{2g})$ 
increases to 1.0 and $n(e_{g})$ decreases to 0.25 eventually, which agree well with the LS configuration of 
Co$^{2+}$ ion ($t^{6}_{2g}e^{1}_{g}$ character). Thus the evolution of Co $3d$ occupancies with respect to 
external pressure provides an strong evidence for the HS-LS spin state crossover in CoO. 

\begin{figure}
\centering
\includegraphics[scale=0.35]{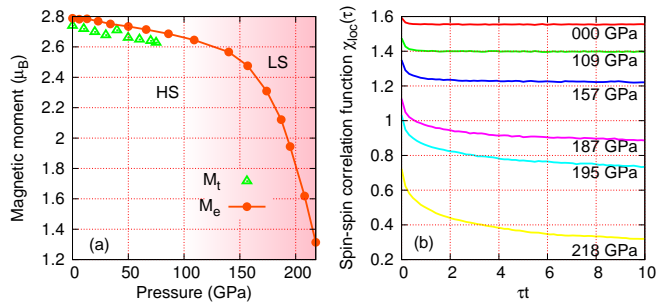}
\caption{(Color online) Spin state transition in CoO. (a) Evolution of the local spin magnetic moment of 
Co $3d$ states. Here $M_e$ means effective local moment which can be calculated by $\sqrt{T\chi_{loc}}$, and 
$M_t$ (illustrated by triangle) denotes the previous theoretical results\cite{wdowik:1698}. (b) Spin-spin 
correlation function $\chi_{loc}(\tau) = \langle S_z (0) S_z (\tau)\rangle $ under various external 
pressures.\label{fig:sz}}
\end{figure}

During the Monte Carlo simulation, we keep track of the different atomic state configurations visited
and draw them as histograms, which give complementary information to the variations of occupancies and 
magnetic states. In Fig.\ref{fig:occ}(b), we show the histograms for several uppermost $ N=7 $ and 
$N=8$ atomic state configurations. It is apparent that at ambient pressure, the HS state ($t^{5}_{2g}e^{2}_{g}$, $S = 3/2$) 
makes predominant contribution and the contribution from LS state ($t^{6}_{2g}e^{1}_{g}$, $S = 1/2$) and 
intermediate spin (IS) state ($t^{6}_{2g}e^{2}_{g}$, $S = 1$) can be ignored reasonably. As increasing the 
pressure, the contributions from HS state have dropped and LS state and IS state tend to grow, and the 
total spin magnetic moment will decrease as well (see Fig.\ref{fig:sz}). We note that except for 
$t^{5}_{2g}e^{2}_{g}$ configuration, the contribution from another HS configuration $t^{4}_{2g}e^{3}_{g}$ ($S = 3/2$) 
is very little. Thus it is confirmed that the HS-LS spin state transition for CoO has a $t^{5}_{2g}e^{2}_{g} 
\rightarrow t^{6}_{2g}e^{1}_{g}$ character, which is consistent with previous assumptions\cite{zhang:155123,rueff:s717}.
In addition, the contribution to magnetic collapse in CoO from HS-IS spin state transition ($t^{5}_{2g}e^{2}_{g} 
\rightarrow t^{6}_{2g}e^{2}_{g}$ character) can not be neglected as well. 

Next, we concentrate our attentions to the magnetic properties of cubic CoO under pressure. In Fig.\ref{fig:sz}(a) it 
shows the evolution of the local spin magnetic moment with pressure. We note that the effective local moment 
$M_e$ is defined through the local spin susceptibility $\sqrt{T \chi_{loc}} $, where $\chi_{loc}$ is defined as 
$\chi_{loc} = \int^{\beta}_{0} d\tau \chi_{loc}(\tau) = \int^{\beta}_{0} d\tau \langle S_z(0) S_z(\tau) \rangle $. 
As is clearly seen in Fig.\ref{fig:sz}(a), under compression the local moment decreases slightly from its 
ambient pressure HS value down to about 170 GPa. Further compression rapidly degrades the moment, which is 
accompanied by the redistribution of electrons $e_{g} \rightarrow t_{2g}$ within the Co $3d$ shell (see Fig.\ref{fig:occ}). 
It should be pointed out that at low pressure region our results coincide with previously published theoretical 
results\cite{wdowik:1698,cohen:654}.

In Fig.\ref{fig:sz}(b), the spin-spin correlation function $\chi_{loc}(\tau)$ under various external pressure 
values are illustrated. For high pressure, a Fermi liquid phase (LS state) is identified. While for low pressure, CoO
exhibits a well-defined frozen moment phase (HS state). In the Fermi liquid phase, the spin-spin correlation 
function behaves as $\chi(\tau) \sim (T/\sin(T\tau\pi))^2$ for times $\tau$ sufficiently far from $\tau = 0$ 
or $\beta$, respectively. For P $\geq 170$ GPa, it displays significant Fermi liquid behaviors, which is 
consistent with the obtained OSMT phase diagram (see Fig.\ref{fig:quasi}) for CoO. The frozen moment 
phase is characterized by a spin-spin correlation function that approaches a non-zero constants at large 
times, as is seen from 0 to 170 GPa. Thus this figure reveals a phase transition between a LS Fermi liquid
metallic phase and a HS Mott phase with frozen moments again.

Now we discuss the most interesting issue of this paper, the relationship between OSMT and HS-LS transition.
The detail analysis of the spin state indicates that the HS-LS transition is the most important driving force
for the OSMT in CoO. Once the Co$^{2+}$ ions keep in HS state, the $e_g$ bands are always half filled, which 
greatly favors Mott insulator phase and very hard to become metallic. On the other hand, in the HS state the 
filling factor of $t_{2g}$ bands is only $1/6$ (in terms of hole density), which makes it much easier to become 
metallic. For example, the previous studies on multiband Hubbard model\cite{han:R4199} show that the critical 
$U_c$ for half-filled two-bands model is around $1.7W$ and that of the $1/6$ filling three-bands model is around 
$2.1W$, where $W$ is the band width. Thus apparently it is the different situation in filling factors between 
$e_g$ and $t_{2g}$ bands which makes the two sub-shells behave so differently under pressure. Once the LS state 
stabilized above 170 GPa, the $e_g$ orbitals are no longer half filling and turn to the metallic phase eventually.

In summary, we conclude that the two-step like Mott transition in CoO can be understood as a typical OSMT. 
At ambient 
pressure CoO is a typical Mott insulator with energy gap around 2.4 eV. At the first transition around 60 GPa 
the $t_{2g}$ bands become metallic while the $e_{g}$ bands still remains insulating until the pressure reaches 
170 GPa. Therefore in CoO the intriguing OSMP with metallic $t_{2g}$ and insulating $e_g$ bands is stable in a 
quite large pressure window between 60 and 170 GPa. Our theoretical calculations also find that the Co$^{2+}$ ions
remain in HS states during the first transition and the crossover to the LS states starts only after the second 
transition, which is in good agreement with both the resistivity\cite{atou:1281} and XES data\cite{rueff:s717,
kurm:165127} for CoO. Further analysis of the calculated results show that the HS-LS transition is the main driving 
force of the Mott MIT in cubic CoO.

\begin{acknowledgments}
We acknowledge financial support from the National Science Foundation ͑of China and that
from the 973 program of China under Contract No.2007CB925000 and No. 2011CBA00108. All the 
LDA+DMFT calculations have been performed on the SHENTENG7000 at Supercomputing Center of Chinese Academy of
Sciences (SCCAS).
\end{acknowledgments}

\bibliography{coo}

\begin{thebibliography}{34}%
\makeatletter
\providecommand \@ifxundefined [1]{%
 \@ifx{#1\undefined}
}%
\providecommand \@ifnum [1]{%
 \ifnum #1\expandafter \@firstoftwo
 \else \expandafter \@secondoftwo
 \fi
}%
\providecommand \@ifx [1]{%
 \ifx #1\expandafter \@firstoftwo
 \else \expandafter \@secondoftwo
 \fi
}%
\providecommand \natexlab [1]{#1}%
\providecommand \enquote  [1]{``#1''}%
\providecommand \bibnamefont  [1]{#1}%
\providecommand \bibfnamefont [1]{#1}%
\providecommand \citenamefont [1]{#1}%
\providecommand \href@noop [0]{\@secondoftwo}%
\providecommand \href [0]{\begingroup \@sanitize@url \@href}%
\providecommand \@href[1]{\@@startlink{#1}\@@href}%
\providecommand \@@href[1]{\endgroup#1\@@endlink}%
\providecommand \@sanitize@url [0]{\catcode `\\12\catcode `\$12\catcode
  `\&12\catcode `\#12\catcode `\^12\catcode `\_12\catcode `\%12\relax}%
\providecommand \@@startlink[1]{}%
\providecommand \@@endlink[0]{}%
\providecommand \url  [0]{\begingroup\@sanitize@url \@url }%
\providecommand \@url [1]{\endgroup\@href {#1}{\urlprefix }}%
\providecommand \urlprefix  [0]{URL }%
\providecommand \Eprint [0]{\href }%
\providecommand \doibase [0]{http://dx.doi.org/}%
\providecommand \selectlanguage [0]{\@gobble}%
\providecommand \bibinfo  [0]{\@secondoftwo}%
\providecommand \bibfield  [0]{\@secondoftwo}%
\providecommand \translation [1]{[#1]}%
\providecommand \BibitemOpen [0]{}%
\providecommand \bibitemStop [0]{}%
\providecommand \bibitemNoStop [0]{.\EOS\space}%
\providecommand \EOS [0]{\spacefactor3000\relax}%
\providecommand \BibitemShut  [1]{\csname bibitem#1\endcsname}%
\let\auto@bib@innerbib\@empty
\bibitem [{\citenamefont {Imada}\ \emph {et~al.}(1998)\citenamefont {Imada},
  \citenamefont {Fujimori},\ and\ \citenamefont {Tokura}}]{imada:1039}%
  \BibitemOpen
  \bibfield  {author} {\bibinfo {author} {\bibfnamefont {M.}~\bibnamefont
  {Imada}}, \bibinfo {author} {\bibfnamefont {A.}~\bibnamefont {Fujimori}}, \
  and\ \bibinfo {author} {\bibfnamefont {Y.}~\bibnamefont {Tokura}},\
  }\href@noop {} {\bibfield  {journal} {\bibinfo  {journal} {Rev. Mod. Phys.}\
  }\textbf {\bibinfo {volume} {70}},\ \bibinfo {pages} {1039} (\bibinfo {year}
  {1998})}\BibitemShut {NoStop}%
\bibitem [{\citenamefont {Anisimov}\ \emph {et~al.}(2002)\citenamefont
  {Anisimov}, \citenamefont {Nekrasov}, \citenamefont {Kondakov}, \citenamefont
  {Rice},\ and\ \citenamefont {Sigrist}}]{anisimov:191}%
  \BibitemOpen
  \bibfield  {author} {\bibinfo {author} {\bibfnamefont {V.~I.}\ \bibnamefont
  {Anisimov}}, \bibinfo {author} {\bibfnamefont {I.~A.}\ \bibnamefont
  {Nekrasov}}, \bibinfo {author} {\bibfnamefont {D.~E.}\ \bibnamefont
  {Kondakov}}, \bibinfo {author} {\bibfnamefont {T.~M.}\ \bibnamefont {Rice}},
  \ and\ \bibinfo {author} {\bibfnamefont {M.}~\bibnamefont {Sigrist}},\
  }\href@noop {} {\bibfield  {journal} {\bibinfo  {journal} {Eur. Phys. J. B}\
  }\textbf {\bibinfo {volume} {25}},\ \bibinfo {pages} {191} (\bibinfo {year}
  {2002})}\BibitemShut {NoStop}%
\bibitem [{\citenamefont {Liebsch}(2003)}]{liebsch:226401}%
  \BibitemOpen
  \bibfield  {author} {\bibinfo {author} {\bibfnamefont {A.}~\bibnamefont
  {Liebsch}},\ }\href@noop {} {\bibfield  {journal} {\bibinfo  {journal} {Phys.
  Rev. Lett.}\ }\textbf {\bibinfo {volume} {91}},\ \bibinfo {pages} {226401}
  (\bibinfo {year} {2003})}\BibitemShut {NoStop}%
\bibitem [{\citenamefont {Koga}\ \emph {et~al.}(2004)\citenamefont {Koga},
  \citenamefont {Kawakami}, \citenamefont {Rice},\ and\ \citenamefont
  {Sigrist}}]{koga:216402}%
  \BibitemOpen
  \bibfield  {author} {\bibinfo {author} {\bibfnamefont {A.}~\bibnamefont
  {Koga}}, \bibinfo {author} {\bibfnamefont {N.}~\bibnamefont {Kawakami}},
  \bibinfo {author} {\bibfnamefont {T.~M.}\ \bibnamefont {Rice}}, \ and\
  \bibinfo {author} {\bibfnamefont {M.}~\bibnamefont {Sigrist}},\ }\href@noop
  {} {\bibfield  {journal} {\bibinfo  {journal} {Phys. Rev. Lett.}\ }\textbf
  {\bibinfo {volume} {92}},\ \bibinfo {pages} {216402} (\bibinfo {year}
  {2004})}\BibitemShut {NoStop}%
\bibitem [{\citenamefont {Koga}\ \emph {et~al.}(2005)\citenamefont {Koga},
  \citenamefont {Kawakami}, \citenamefont {Rice},\ and\ \citenamefont
  {Sigrist}}]{koga:045128}%
  \BibitemOpen
  \bibfield  {author} {\bibinfo {author} {\bibfnamefont {A.}~\bibnamefont
  {Koga}}, \bibinfo {author} {\bibfnamefont {N.}~\bibnamefont {Kawakami}},
  \bibinfo {author} {\bibfnamefont {T.~M.}\ \bibnamefont {Rice}}, \ and\
  \bibinfo {author} {\bibfnamefont {M.}~\bibnamefont {Sigrist}},\ }\href@noop
  {} {\bibfield  {journal} {\bibinfo  {journal} {Phys. Rev. B}\ }\textbf
  {\bibinfo {volume} {72}},\ \bibinfo {pages} {045128} (\bibinfo {year}
  {2005})}\BibitemShut {NoStop}%
\bibitem [{\citenamefont {de'Medici}\ \emph {et~al.}(2005)\citenamefont
  {de'Medici}, \citenamefont {Georges},\ and\ \citenamefont
  {Biermann}}]{de'medici:205124}%
  \BibitemOpen
  \bibfield  {author} {\bibinfo {author} {\bibfnamefont {L.}~\bibnamefont
  {de'Medici}}, \bibinfo {author} {\bibfnamefont {A.}~\bibnamefont {Georges}},
  \ and\ \bibinfo {author} {\bibfnamefont {S.}~\bibnamefont {Biermann}},\
  }\href@noop {} {\bibfield  {journal} {\bibinfo  {journal} {Phys. Rev. B}\
  }\textbf {\bibinfo {volume} {72}},\ \bibinfo {pages} {205124} (\bibinfo
  {year} {2005})}\BibitemShut {NoStop}%
\bibitem [{\citenamefont {de' Medici}\ \emph {et~al.}(2009)\citenamefont {de'
  Medici}, \citenamefont {Hassan}, \citenamefont {Capone},\ and\ \citenamefont
  {Dai}}]{medici:126401}%
  \BibitemOpen
  \bibfield  {author} {\bibinfo {author} {\bibfnamefont {L.}~\bibnamefont {de'
  Medici}}, \bibinfo {author} {\bibfnamefont {S.~R.}\ \bibnamefont {Hassan}},
  \bibinfo {author} {\bibfnamefont {M.}~\bibnamefont {Capone}}, \ and\ \bibinfo
  {author} {\bibfnamefont {X.}~\bibnamefont {Dai}},\ }\href@noop {} {\bibfield
  {journal} {\bibinfo  {journal} {Phys. Rev. Lett.}\ }\textbf {\bibinfo
  {volume} {102}},\ \bibinfo {pages} {126401} (\bibinfo {year}
  {2009})}\BibitemShut {NoStop}%
\bibitem [{\citenamefont {Werner}\ and\ \citenamefont
  {Millis}(2007)}]{werner:126405}%
  \BibitemOpen
  \bibfield  {author} {\bibinfo {author} {\bibfnamefont {P.}~\bibnamefont
  {Werner}}\ and\ \bibinfo {author} {\bibfnamefont {A.~J.}\ \bibnamefont
  {Millis}},\ }\href@noop {} {\bibfield  {journal} {\bibinfo  {journal} {Phys.
  Rev. Lett.}\ }\textbf {\bibinfo {volume} {99}},\ \bibinfo {pages} {126405}
  (\bibinfo {year} {2007})}\BibitemShut {NoStop}%
\bibitem [{\citenamefont {Kune\v{s}}\ \emph {et~al.}(2008)\citenamefont
  {Kune\v{s}}, \citenamefont {Lukoyanov}, \citenamefont {Anisimov},
  \citenamefont {Scalettar},\ and\ \citenamefont {Pickett}}]{kunes:2008}%
  \BibitemOpen
  \bibfield  {author} {\bibinfo {author} {\bibfnamefont {J.}~\bibnamefont
  {Kune\v{s}}}, \bibinfo {author} {\bibfnamefont {A.~V.}\ \bibnamefont
  {Lukoyanov}}, \bibinfo {author} {\bibfnamefont {V.~I.}\ \bibnamefont
  {Anisimov}}, \bibinfo {author} {\bibfnamefont {R.~T.}\ \bibnamefont
  {Scalettar}}, \ and\ \bibinfo {author} {\bibfnamefont {W.~E.}\ \bibnamefont
  {Pickett}},\ }\href@noop {} {\bibfield  {journal} {\bibinfo  {journal} {Nat
  Mater}\ }\textbf {\bibinfo {volume} {7}},\ \bibinfo {pages} {198} (\bibinfo
  {year} {2008})}\BibitemShut {NoStop}%
\bibitem [{\citenamefont {Kune\v{s}}\ \emph {et~al.}(2009)\citenamefont
  {Kune\v{s}}, \citenamefont {Korotin}, \citenamefont {Korotin}, \citenamefont
  {Anisimov},\ and\ \citenamefont {Werner}}]{kunes:146402}%
  \BibitemOpen
  \bibfield  {author} {\bibinfo {author} {\bibfnamefont {J.}~\bibnamefont
  {Kune\v{s}}}, \bibinfo {author} {\bibfnamefont {D.~M.}\ \bibnamefont
  {Korotin}}, \bibinfo {author} {\bibfnamefont {M.~A.}\ \bibnamefont
  {Korotin}}, \bibinfo {author} {\bibfnamefont {V.~I.}\ \bibnamefont
  {Anisimov}}, \ and\ \bibinfo {author} {\bibfnamefont {P.}~\bibnamefont
  {Werner}},\ }\href@noop {} {\bibfield  {journal} {\bibinfo  {journal} {Phys.
  Rev. Lett.}\ }\textbf {\bibinfo {volume} {102}},\ \bibinfo {pages} {146402}
  (\bibinfo {year} {2009})}\BibitemShut {NoStop}%
\bibitem [{\citenamefont {Shorikov}\ \emph {et~al.}(2010)\citenamefont
  {Shorikov}, \citenamefont {Pchelkina}, \citenamefont {Anisimov},
  \citenamefont {Skornyakov},\ and\ \citenamefont {Korotin}}]{shorikov:195101}%
  \BibitemOpen
  \bibfield  {author} {\bibinfo {author} {\bibfnamefont {A.~O.}\ \bibnamefont
  {Shorikov}}, \bibinfo {author} {\bibfnamefont {Z.~V.}\ \bibnamefont
  {Pchelkina}}, \bibinfo {author} {\bibfnamefont {V.~I.}\ \bibnamefont
  {Anisimov}}, \bibinfo {author} {\bibfnamefont {S.~L.}\ \bibnamefont
  {Skornyakov}}, \ and\ \bibinfo {author} {\bibfnamefont {M.~A.}\ \bibnamefont
  {Korotin}},\ }\href@noop {} {\bibfield  {journal} {\bibinfo  {journal} {Phys.
  Rev. B}\ }\textbf {\bibinfo {volume} {82}},\ \bibinfo {pages} {195101}
  (\bibinfo {year} {2010})}\BibitemShut {NoStop}%
\bibitem [{\citenamefont {Lyubutin}\ \emph {et~al.}(2009)\citenamefont
  {Lyubutin}, \citenamefont {Ovchinnikov}, \citenamefont {Gavriliuk},\ and\
  \citenamefont {Struzhkin}}]{lyub:085125}%
  \BibitemOpen
  \bibfield  {author} {\bibinfo {author} {\bibfnamefont {I.~S.}\ \bibnamefont
  {Lyubutin}}, \bibinfo {author} {\bibfnamefont {S.~G.}\ \bibnamefont
  {Ovchinnikov}}, \bibinfo {author} {\bibfnamefont {A.~G.}\ \bibnamefont
  {Gavriliuk}}, \ and\ \bibinfo {author} {\bibfnamefont {V.~V.}\ \bibnamefont
  {Struzhkin}},\ }\href@noop {} {\bibfield  {journal} {\bibinfo  {journal}
  {Phys. Rev. B}\ }\textbf {\bibinfo {volume} {79}},\ \bibinfo {pages} {085125}
  (\bibinfo {year} {2009})}\BibitemShut {NoStop}%
\bibitem [{\citenamefont {Cohen}\ \emph {et~al.}(1997)\citenamefont {Cohen},
  \citenamefont {Mazin},\ and\ \citenamefont {Isaak}}]{cohen:654}%
  \BibitemOpen
  \bibfield  {author} {\bibinfo {author} {\bibfnamefont {R.~E.}\ \bibnamefont
  {Cohen}}, \bibinfo {author} {\bibfnamefont {I.~I.}\ \bibnamefont {Mazin}}, \
  and\ \bibinfo {author} {\bibfnamefont {D.~G.}\ \bibnamefont {Isaak}},\
  }\href@noop {} {\bibfield  {journal} {\bibinfo  {journal} {Science}\ }\textbf
  {\bibinfo {volume} {275}},\ \bibinfo {pages} {654} (\bibinfo {year}
  {1997})}\BibitemShut {NoStop}%
\bibitem [{\citenamefont {Gavriliuk}\ \emph {et~al.}(2008)\citenamefont
  {Gavriliuk}, \citenamefont {Struzhkin}, \citenamefont {Lyubutin},
  \citenamefont {Ovchinnikov}, \citenamefont {Hu},\ and\ \citenamefont
  {Chow}}]{gavr:155112}%
  \BibitemOpen
  \bibfield  {author} {\bibinfo {author} {\bibfnamefont {A.~G.}\ \bibnamefont
  {Gavriliuk}}, \bibinfo {author} {\bibfnamefont {V.~V.}\ \bibnamefont
  {Struzhkin}}, \bibinfo {author} {\bibfnamefont {I.~S.}\ \bibnamefont
  {Lyubutin}}, \bibinfo {author} {\bibfnamefont {S.~G.}\ \bibnamefont
  {Ovchinnikov}}, \bibinfo {author} {\bibfnamefont {M.~Y.}\ \bibnamefont {Hu}},
  \ and\ \bibinfo {author} {\bibfnamefont {P.}~\bibnamefont {Chow}},\
  }\href@noop {} {\bibfield  {journal} {\bibinfo  {journal} {Phys. Rev. B}\
  }\textbf {\bibinfo {volume} {77}},\ \bibinfo {pages} {155112} (\bibinfo
  {year} {2008})}\BibitemShut {NoStop}%
\bibitem [{\citenamefont {Mattila}\ \emph {et~al.}(2007)\citenamefont
  {Mattila}, \citenamefont {Rueff}, \citenamefont {Badro}, \citenamefont
  {Vank\'o},\ and\ \citenamefont {Shukla}}]{mattila:196404}%
  \BibitemOpen
  \bibfield  {author} {\bibinfo {author} {\bibfnamefont {A.}~\bibnamefont
  {Mattila}}, \bibinfo {author} {\bibfnamefont {J.-P.}\ \bibnamefont {Rueff}},
  \bibinfo {author} {\bibfnamefont {J.}~\bibnamefont {Badro}}, \bibinfo
  {author} {\bibfnamefont {G.}~\bibnamefont {Vank\'o}}, \ and\ \bibinfo
  {author} {\bibfnamefont {A.}~\bibnamefont {Shukla}},\ }\href@noop {}
  {\bibfield  {journal} {\bibinfo  {journal} {Phys. Rev. Lett.}\ }\textbf
  {\bibinfo {volume} {98}},\ \bibinfo {pages} {196404} (\bibinfo {year}
  {2007})}\BibitemShut {NoStop}%
\bibitem [{\citenamefont {Atou}\ \emph {et~al.}(2004)\citenamefont {Atou},
  \citenamefont {Kawasaki},\ and\ \citenamefont {Nakajima}}]{atou:1281}%
  \BibitemOpen
  \bibfield  {author} {\bibinfo {author} {\bibfnamefont {T.}~\bibnamefont
  {Atou}}, \bibinfo {author} {\bibfnamefont {M.}~\bibnamefont {Kawasaki}}, \
  and\ \bibinfo {author} {\bibfnamefont {S.}~\bibnamefont {Nakajima}},\
  }\href@noop {} {\bibfield  {journal} {\bibinfo  {journal} {Jpn. J. Appl.
  Phys.}\ }\textbf {\bibinfo {volume} {43}},\ \bibinfo {pages} {L1281}
  (\bibinfo {year} {2004})}\BibitemShut {NoStop}%
\bibitem [{\citenamefont {Rueff}\ \emph {et~al.}(2005)\citenamefont {Rueff},
  \citenamefont {Mattila}, \citenamefont {Badro}, \citenamefont {Vankó},\ and\
  \citenamefont {Shukla}}]{rueff:s717}%
  \BibitemOpen
  \bibfield  {author} {\bibinfo {author} {\bibfnamefont {J.-P.}\ \bibnamefont
  {Rueff}}, \bibinfo {author} {\bibfnamefont {A.}~\bibnamefont {Mattila}},
  \bibinfo {author} {\bibfnamefont {J.}~\bibnamefont {Badro}}, \bibinfo
  {author} {\bibfnamefont {G.}~\bibnamefont {Vankó}}, \ and\ \bibinfo {author}
  {\bibfnamefont {A.}~\bibnamefont {Shukla}},\ }\href@noop {} {\bibfield
  {journal} {\bibinfo  {journal} {J. Phys.: Condens. Matter}\ }\textbf
  {\bibinfo {volume} {17}},\ \bibinfo {pages} {S717} (\bibinfo {year}
  {2005})}\BibitemShut {NoStop}%
\bibitem [{\citenamefont {Kurmaev}\ \emph {et~al.}(2008)\citenamefont
  {Kurmaev}, \citenamefont {Wilks}, \citenamefont {Moewes}, \citenamefont
  {Finkelstein}, \citenamefont {Shamin},\ and\ \citenamefont
  {Kune\v{s}}}]{kurm:165127}%
  \BibitemOpen
  \bibfield  {author} {\bibinfo {author} {\bibfnamefont {E.~Z.}\ \bibnamefont
  {Kurmaev}}, \bibinfo {author} {\bibfnamefont {R.~G.}\ \bibnamefont {Wilks}},
  \bibinfo {author} {\bibfnamefont {A.}~\bibnamefont {Moewes}}, \bibinfo
  {author} {\bibfnamefont {L.~D.}\ \bibnamefont {Finkelstein}}, \bibinfo
  {author} {\bibfnamefont {S.~N.}\ \bibnamefont {Shamin}}, \ and\ \bibinfo
  {author} {\bibfnamefont {J.}~\bibnamefont {Kune\v{s}}},\ }\href@noop {}
  {\bibfield  {journal} {\bibinfo  {journal} {Phys. Rev. B}\ }\textbf {\bibinfo
  {volume} {77}},\ \bibinfo {pages} {165127} (\bibinfo {year}
  {2008})}\BibitemShut {NoStop}%
\bibitem [{\citenamefont {Kasinathan}\ \emph {et~al.}(2006)\citenamefont
  {Kasinathan}, \citenamefont {Kune\v{s}}, \citenamefont {Koepernik},
  \citenamefont {Diaconu}, \citenamefont {Martin}, \citenamefont {Prodan},
  \citenamefont {Scuseria}, \citenamefont {Spaldin}, \citenamefont {Petit},
  \citenamefont {Schulthess},\ and\ \citenamefont {Pickett}}]{kas:195110}%
  \BibitemOpen
  \bibfield  {author} {\bibinfo {author} {\bibfnamefont {D.}~\bibnamefont
  {Kasinathan}}, \bibinfo {author} {\bibfnamefont {J.}~\bibnamefont
  {Kune\v{s}}}, \bibinfo {author} {\bibfnamefont {K.}~\bibnamefont
  {Koepernik}}, \bibinfo {author} {\bibfnamefont {C.~V.}\ \bibnamefont
  {Diaconu}}, \bibinfo {author} {\bibfnamefont {R.~L.}\ \bibnamefont {Martin}},
  \bibinfo {author} {\bibfnamefont {I.~D.}\ \bibnamefont {Prodan}}, \bibinfo
  {author} {\bibfnamefont {G.~E.}\ \bibnamefont {Scuseria}}, \bibinfo {author}
  {\bibfnamefont {N.}~\bibnamefont {Spaldin}}, \bibinfo {author} {\bibfnamefont
  {L.}~\bibnamefont {Petit}}, \bibinfo {author} {\bibfnamefont {T.~C.}\
  \bibnamefont {Schulthess}}, \ and\ \bibinfo {author} {\bibfnamefont {W.~E.}\
  \bibnamefont {Pickett}},\ }\href@noop {} {\bibfield  {journal} {\bibinfo
  {journal} {Phys. Rev. B}\ }\textbf {\bibinfo {volume} {74}},\ \bibinfo
  {pages} {195110} (\bibinfo {year} {2006})}\BibitemShut {NoStop}%
\bibitem [{\citenamefont {Wdowik}\ and\ \citenamefont
  {Legut}(2008)}]{wdowik:1698}%
  \BibitemOpen
  \bibfield  {author} {\bibinfo {author} {\bibfnamefont {U.}~\bibnamefont
  {Wdowik}}\ and\ \bibinfo {author} {\bibfnamefont {D.}~\bibnamefont {Legut}},\
  }\href@noop {} {\bibfield  {journal} {\bibinfo  {journal} {J. Phys. Chem.
  Sol.}\ }\textbf {\bibinfo {volume} {69}},\ \bibinfo {pages} {1698} (\bibinfo
  {year} {2008})}\BibitemShut {NoStop}%
\bibitem [{\citenamefont {Zhang}\ \emph {et~al.}(2009)\citenamefont {Zhang},
  \citenamefont {Koepernik}, \citenamefont {Richter},\ and\ \citenamefont
  {Eschrig}}]{zhang:155123}%
  \BibitemOpen
  \bibfield  {author} {\bibinfo {author} {\bibfnamefont {W.}~\bibnamefont
  {Zhang}}, \bibinfo {author} {\bibfnamefont {K.}~\bibnamefont {Koepernik}},
  \bibinfo {author} {\bibfnamefont {M.}~\bibnamefont {Richter}}, \ and\
  \bibinfo {author} {\bibfnamefont {H.}~\bibnamefont {Eschrig}},\ }\href@noop
  {} {\bibfield  {journal} {\bibinfo  {journal} {Phys. Rev. B}\ }\textbf
  {\bibinfo {volume} {79}},\ \bibinfo {pages} {155123} (\bibinfo {year}
  {2009})}\BibitemShut {NoStop}%
\bibitem [{\citenamefont {Georges}\ \emph {et~al.}(1996)\citenamefont
  {Georges}, \citenamefont {Kotliar}, \citenamefont {Krauth},\ and\
  \citenamefont {Rozenberg}}]{antoine:13}%
  \BibitemOpen
  \bibfield  {author} {\bibinfo {author} {\bibfnamefont {A.}~\bibnamefont
  {Georges}}, \bibinfo {author} {\bibfnamefont {G.}~\bibnamefont {Kotliar}},
  \bibinfo {author} {\bibfnamefont {W.}~\bibnamefont {Krauth}}, \ and\ \bibinfo
  {author} {\bibfnamefont {M.~J.}\ \bibnamefont {Rozenberg}},\ }\href@noop {}
  {\bibfield  {journal} {\bibinfo  {journal} {Rev. Mod. Phys.}\ }\textbf
  {\bibinfo {volume} {68}},\ \bibinfo {pages} {13} (\bibinfo {year}
  {1996})}\BibitemShut {NoStop}%
\bibitem [{\citenamefont {Kotliar}\ \emph {et~al.}(2006)\citenamefont
  {Kotliar}, \citenamefont {Savrasov}, \citenamefont {Haule}, \citenamefont
  {Oudovenko}, \citenamefont {Parcollet},\ and\ \citenamefont
  {Marianetti}}]{kotliar:865}%
  \BibitemOpen
  \bibfield  {author} {\bibinfo {author} {\bibfnamefont {G.}~\bibnamefont
  {Kotliar}}, \bibinfo {author} {\bibfnamefont {S.~Y.}\ \bibnamefont
  {Savrasov}}, \bibinfo {author} {\bibfnamefont {K.}~\bibnamefont {Haule}},
  \bibinfo {author} {\bibfnamefont {V.~S.}\ \bibnamefont {Oudovenko}}, \bibinfo
  {author} {\bibfnamefont {O.}~\bibnamefont {Parcollet}}, \ and\ \bibinfo
  {author} {\bibfnamefont {C.~A.}\ \bibnamefont {Marianetti}},\ }\href@noop {}
  {\bibfield  {journal} {\bibinfo  {journal} {Rev. Mod. Phys.}\ }\textbf
  {\bibinfo {volume} {78}},\ \bibinfo {pages} {865} (\bibinfo {year}
  {2006})}\BibitemShut {NoStop}%
\bibitem [{\citenamefont {Korotin}\ \emph {et~al.}(2008)\citenamefont
  {Korotin}, \citenamefont {Kozhevnikov}, \citenamefont {Skornyakov},
  \citenamefont {Leonov}, \citenamefont {Binggeli}, \citenamefont {Anisimov},\
  and\ \citenamefont {Trimarchi}}]{korotin:91}%
  \BibitemOpen
  \bibfield  {author} {\bibinfo {author} {\bibfnamefont {D.}~\bibnamefont
  {Korotin}}, \bibinfo {author} {\bibfnamefont {A.}~\bibnamefont
  {Kozhevnikov}}, \bibinfo {author} {\bibfnamefont {S.}~\bibnamefont
  {Skornyakov}}, \bibinfo {author} {\bibfnamefont {I.}~\bibnamefont {Leonov}},
  \bibinfo {author} {\bibfnamefont {N.}~\bibnamefont {Binggeli}}, \bibinfo
  {author} {\bibfnamefont {V.}~\bibnamefont {Anisimov}}, \ and\ \bibinfo
  {author} {\bibfnamefont {G.}~\bibnamefont {Trimarchi}},\ }\href@noop {}
  {\bibfield  {journal} {\bibinfo  {journal} {Eur. Phys. J. B}\ }\textbf
  {\bibinfo {volume} {65}},\ \bibinfo {pages} {91} (\bibinfo {year}
  {2008})}\BibitemShut {NoStop}%
\bibitem [{\citenamefont {Trimarchi}\ \emph {et~al.}(2008)\citenamefont
  {Trimarchi}, \citenamefont {Leonov}, \citenamefont {Binggeli}, \citenamefont
  {Korotin},\ and\ \citenamefont {Anisimov}}]{trimarchi:135227}%
  \BibitemOpen
  \bibfield  {author} {\bibinfo {author} {\bibfnamefont {G.}~\bibnamefont
  {Trimarchi}}, \bibinfo {author} {\bibfnamefont {I.}~\bibnamefont {Leonov}},
  \bibinfo {author} {\bibfnamefont {N.}~\bibnamefont {Binggeli}}, \bibinfo
  {author} {\bibfnamefont {D.}~\bibnamefont {Korotin}}, \ and\ \bibinfo
  {author} {\bibfnamefont {V.~I.}\ \bibnamefont {Anisimov}},\ }\href@noop {}
  {\bibfield  {journal} {\bibinfo  {journal} {J. Phys: Condens. Matter}\
  }\textbf {\bibinfo {volume} {20}},\ \bibinfo {pages} {135227} (\bibinfo
  {year} {2008})}\BibitemShut {NoStop}%
\bibitem [{\citenamefont {Amadon}\ \emph {et~al.}(2008)\citenamefont {Amadon},
  \citenamefont {Lechermann}, \citenamefont {Georges}, \citenamefont {Jollet},
  \citenamefont {Wehling},\ and\ \citenamefont {Lichtenstein}}]{amadon:205112}%
  \BibitemOpen
  \bibfield  {author} {\bibinfo {author} {\bibfnamefont {B.}~\bibnamefont
  {Amadon}}, \bibinfo {author} {\bibfnamefont {F.}~\bibnamefont {Lechermann}},
  \bibinfo {author} {\bibfnamefont {A.}~\bibnamefont {Georges}}, \bibinfo
  {author} {\bibfnamefont {F.}~\bibnamefont {Jollet}}, \bibinfo {author}
  {\bibfnamefont {T.~O.}\ \bibnamefont {Wehling}}, \ and\ \bibinfo {author}
  {\bibfnamefont {A.~I.}\ \bibnamefont {Lichtenstein}},\ }\href@noop {}
  {\bibfield  {journal} {\bibinfo  {journal} {Phys. Rev. B}\ }\textbf {\bibinfo
  {volume} {77}},\ \bibinfo {pages} {205112} (\bibinfo {year}
  {2008})}\BibitemShut {NoStop}%
\bibitem [{\citenamefont {Giannozzi}\ \emph {et~al.}(2009)\citenamefont
  {Giannozzi}, \citenamefont {Baroni}, \citenamefont {Bonini}, \citenamefont
  {Calandra}, \citenamefont {Car}, \citenamefont {Cavazzoni}, \citenamefont
  {Ceresoli}, \citenamefont {Chiarotti}, \citenamefont {Cococcioni},
  \citenamefont {Dabo}, \citenamefont {{Dal Corso}}, \citenamefont
  {de~Gironcoli}, \citenamefont {Fabris}, \citenamefont {Fratesi},
  \citenamefont {Gebauer}, \citenamefont {Gerstmann}, \citenamefont
  {Gougoussis}, \citenamefont {Kokalj}, \citenamefont {Lazzeri}, \citenamefont
  {Martin-Samos}, \citenamefont {Marzari}, \citenamefont {Mauri}, \citenamefont
  {Mazzarello}, \citenamefont {Paolini}, \citenamefont {Pasquarello},
  \citenamefont {Paulatto}, \citenamefont {Sbraccia}, \citenamefont {Scandolo},
  \citenamefont {Sclauzero}, \citenamefont {Seitsonen}, \citenamefont
  {Smogunov}, \citenamefont {Umari},\ and\ \citenamefont
  {Wentzcovitch}}]{qe2009}%
  \BibitemOpen
  \bibfield  {author} {\bibinfo {author} {\bibfnamefont {P.}~\bibnamefont
  {Giannozzi}}, \bibinfo {author} {\bibfnamefont {S.}~\bibnamefont {Baroni}},
  \bibinfo {author} {\bibfnamefont {N.}~\bibnamefont {Bonini}}, \bibinfo
  {author} {\bibfnamefont {M.}~\bibnamefont {Calandra}}, \bibinfo {author}
  {\bibfnamefont {R.}~\bibnamefont {Car}}, \bibinfo {author} {\bibfnamefont
  {C.}~\bibnamefont {Cavazzoni}}, \bibinfo {author} {\bibfnamefont
  {D.}~\bibnamefont {Ceresoli}}, \bibinfo {author} {\bibfnamefont {G.~L.}\
  \bibnamefont {Chiarotti}}, \bibinfo {author} {\bibfnamefont {M.}~\bibnamefont
  {Cococcioni}}, \bibinfo {author} {\bibfnamefont {I.}~\bibnamefont {Dabo}},
  \bibinfo {author} {\bibfnamefont {A.}~\bibnamefont {{Dal Corso}}}, \bibinfo
  {author} {\bibfnamefont {S.}~\bibnamefont {de~Gironcoli}}, \bibinfo {author}
  {\bibfnamefont {S.}~\bibnamefont {Fabris}}, \bibinfo {author} {\bibfnamefont
  {G.}~\bibnamefont {Fratesi}}, \bibinfo {author} {\bibfnamefont
  {R.}~\bibnamefont {Gebauer}}, \bibinfo {author} {\bibfnamefont
  {U.}~\bibnamefont {Gerstmann}}, \bibinfo {author} {\bibfnamefont
  {C.}~\bibnamefont {Gougoussis}}, \bibinfo {author} {\bibfnamefont
  {A.}~\bibnamefont {Kokalj}}, \bibinfo {author} {\bibfnamefont
  {M.}~\bibnamefont {Lazzeri}}, \bibinfo {author} {\bibfnamefont
  {L.}~\bibnamefont {Martin-Samos}}, \bibinfo {author} {\bibfnamefont
  {N.}~\bibnamefont {Marzari}}, \bibinfo {author} {\bibfnamefont
  {F.}~\bibnamefont {Mauri}}, \bibinfo {author} {\bibfnamefont
  {R.}~\bibnamefont {Mazzarello}}, \bibinfo {author} {\bibfnamefont
  {S.}~\bibnamefont {Paolini}}, \bibinfo {author} {\bibfnamefont
  {A.}~\bibnamefont {Pasquarello}}, \bibinfo {author} {\bibfnamefont
  {L.}~\bibnamefont {Paulatto}}, \bibinfo {author} {\bibfnamefont
  {C.}~\bibnamefont {Sbraccia}}, \bibinfo {author} {\bibfnamefont
  {S.}~\bibnamefont {Scandolo}}, \bibinfo {author} {\bibfnamefont
  {G.}~\bibnamefont {Sclauzero}}, \bibinfo {author} {\bibfnamefont {A.~P.}\
  \bibnamefont {Seitsonen}}, \bibinfo {author} {\bibfnamefont {A.}~\bibnamefont
  {Smogunov}}, \bibinfo {author} {\bibfnamefont {P.}~\bibnamefont {Umari}}, \
  and\ \bibinfo {author} {\bibfnamefont {R.~M.}\ \bibnamefont {Wentzcovitch}},\
  }\href@noop {} {\bibfield  {journal} {\bibinfo  {journal} {J. Phys: Condens.
  Matter}\ }\textbf {\bibinfo {volume} {21}},\ \bibinfo {pages} {395502}
  (\bibinfo {year} {2009})}\BibitemShut {NoStop}%
\bibitem [{\citenamefont {Werner}\ \emph {et~al.}(2006)\citenamefont {Werner},
  \citenamefont {Comanac}, \citenamefont {de' Medici}, \citenamefont {Troyer},\
  and\ \citenamefont {Millis}}]{werner:076405}%
  \BibitemOpen
  \bibfield  {author} {\bibinfo {author} {\bibfnamefont {P.}~\bibnamefont
  {Werner}}, \bibinfo {author} {\bibfnamefont {A.}~\bibnamefont {Comanac}},
  \bibinfo {author} {\bibfnamefont {L.}~\bibnamefont {de' Medici}}, \bibinfo
  {author} {\bibfnamefont {M.}~\bibnamefont {Troyer}}, \ and\ \bibinfo {author}
  {\bibfnamefont {A.~J.}\ \bibnamefont {Millis}},\ }\href@noop {} {\bibfield
  {journal} {\bibinfo  {journal} {Phys. Rev. Lett.}\ }\textbf {\bibinfo
  {volume} {97}},\ \bibinfo {pages} {076405} (\bibinfo {year}
  {2006})}\BibitemShut {NoStop}%
\bibitem [{\citenamefont {Gull}\ \emph {et~al.}(2011)\citenamefont {Gull},
  \citenamefont {Millis}, \citenamefont {Lichtenstein}, \citenamefont
  {Rubtsov}, \citenamefont {Troyer},\ and\ \citenamefont {Werner}}]{gull:349}%
  \BibitemOpen
  \bibfield  {author} {\bibinfo {author} {\bibfnamefont {E.}~\bibnamefont
  {Gull}}, \bibinfo {author} {\bibfnamefont {A.~J.}\ \bibnamefont {Millis}},
  \bibinfo {author} {\bibfnamefont {A.~I.}\ \bibnamefont {Lichtenstein}},
  \bibinfo {author} {\bibfnamefont {A.~N.}\ \bibnamefont {Rubtsov}}, \bibinfo
  {author} {\bibfnamefont {M.}~\bibnamefont {Troyer}}, \ and\ \bibinfo {author}
  {\bibfnamefont {P.}~\bibnamefont {Werner}},\ }\href@noop {} {\bibfield
  {journal} {\bibinfo  {journal} {Rev. Mod. Phys.}\ }\textbf {\bibinfo {volume}
  {83}},\ \bibinfo {pages} {349} (\bibinfo {year} {2011})}\BibitemShut
  {NoStop}%
\bibitem [{\citenamefont {Anisimov}\ \emph {et~al.}(1991)\citenamefont
  {Anisimov}, \citenamefont {Zaanen},\ and\ \citenamefont
  {Andersen}}]{anis:943}%
  \BibitemOpen
  \bibfield  {author} {\bibinfo {author} {\bibfnamefont {V.~I.}\ \bibnamefont
  {Anisimov}}, \bibinfo {author} {\bibfnamefont {J.}~\bibnamefont {Zaanen}}, \
  and\ \bibinfo {author} {\bibfnamefont {O.~K.}\ \bibnamefont {Andersen}},\
  }\href@noop {} {\bibfield  {journal} {\bibinfo  {journal} {Phys. Rev. B}\
  }\textbf {\bibinfo {volume} {44}},\ \bibinfo {pages} {943} (\bibinfo {year}
  {1991})}\BibitemShut {NoStop}%
\bibitem [{\citenamefont {Boehnke}\ \emph {et~al.}(2011)\citenamefont
  {Boehnke}, \citenamefont {Hafermann}, \citenamefont {Ferrero}, \citenamefont
  {Lechermann},\ and\ \citenamefont {Parcollet}}]{boe:075145}%
  \BibitemOpen
  \bibfield  {author} {\bibinfo {author} {\bibfnamefont {L.}~\bibnamefont
  {Boehnke}}, \bibinfo {author} {\bibfnamefont {H.}~\bibnamefont {Hafermann}},
  \bibinfo {author} {\bibfnamefont {M.}~\bibnamefont {Ferrero}}, \bibinfo
  {author} {\bibfnamefont {F.}~\bibnamefont {Lechermann}}, \ and\ \bibinfo
  {author} {\bibfnamefont {O.}~\bibnamefont {Parcollet}},\ }\href@noop {}
  {\bibfield  {journal} {\bibinfo  {journal} {Phys. Rev. B}\ }\textbf {\bibinfo
  {volume} {84}},\ \bibinfo {pages} {075145} (\bibinfo {year}
  {2011})}\BibitemShut {NoStop}%
\bibitem [{\citenamefont {Jarrell}\ and\ \citenamefont
  {Gubernatis}(1996)}]{jarrell:133}%
  \BibitemOpen
  \bibfield  {author} {\bibinfo {author} {\bibfnamefont {M.}~\bibnamefont
  {Jarrell}}\ and\ \bibinfo {author} {\bibfnamefont {J.}~\bibnamefont
  {Gubernatis}},\ }\href@noop {} {\bibfield  {journal} {\bibinfo  {journal}
  {Phys. Rep.}\ }\textbf {\bibinfo {volume} {269}},\ \bibinfo {pages} {133 }
  (\bibinfo {year} {1996})}\BibitemShut {NoStop}%
\bibitem [{\citenamefont {Beach}()}]{beach}%
  \BibitemOpen
  \bibfield  {author} {\bibinfo {author} {\bibfnamefont {K.~S.~D.}\
  \bibnamefont {Beach}},\ }\href@noop {} {\ }\Eprint
  {http://arxiv.org/abs/0403055} {arXiv:0403055 [cond-mat]} \BibitemShut
  {NoStop}%
\bibitem [{\citenamefont {Han}\ \emph {et~al.}(1998)\citenamefont {Han},
  \citenamefont {Jarrell},\ and\ \citenamefont {Cox}}]{han:R4199}%
  \BibitemOpen
  \bibfield  {author} {\bibinfo {author} {\bibfnamefont {J.~E.}\ \bibnamefont
  {Han}}, \bibinfo {author} {\bibfnamefont {M.}~\bibnamefont {Jarrell}}, \ and\
  \bibinfo {author} {\bibfnamefont {D.~L.}\ \bibnamefont {Cox}},\ }\href@noop
  {} {\bibfield  {journal} {\bibinfo  {journal} {Phys. Rev. B}\ }\textbf
  {\bibinfo {volume} {58}},\ \bibinfo {pages} {R4199} (\bibinfo {year}
  {1998})}\BibitemShut {NoStop}%
\end{thebibliography}%

\end{document}